\documentclass[pre,aps,twocolumn,showpacs, showkeys,floatfix,nofootinbib,superscriptaddress]{revtex4-1}
\usepackage{amsmath}
\usepackage{amsfonts}
\usepackage{amssymb}
\usepackage{graphicx}
\usepackage{subfigure}
\usepackage{float}
\usepackage{ulem}
\usepackage{hyperref}
\usepackage{color}
\usepackage{tikz}
\usetikzlibrary{decorations.pathmorphing,patterns,arrows,shapes.misc}
\usepackage{tikz-3dplot}
\usepackage[nomessages]{fp}
\usepackage[first=0, last=360]{lcg}
\usepackage{ifthen}
\usepackage{csvsimple}
\usepackage{comment}
\begin{document}

\title{Contact Process on Weighted Planar Stochastic Lattice}

\author{Sidiney G. Alves}\email{$^1$sidiney@ufsj.edu.br}
\author{Marcelo M. de Oliveira}\email{mmdeoliveira@ufsj.edu.br}
\address{Departamento de Estat\'istica, F\'isica e Matem\'atica, Universidade Federal de S\~ao Jo\~ao Del-Rei \\
	36490-972, Ouro Branco, MG, Brazil}

\begin{abstract}
We study the absorbing state phase transition in the contact process on the Weighted Planar Stochastic (WPS) Lattice. The WPS lattice is multifractal. Its dual network has a power-law degree distribution function and is also embedded in a bidimensional space. Moreover, it represents a novel way to introduce coordination disorder in lattice models. We investigated the critical behavior of the disordered system using extensive simulations. Our results show the critical behavior is distinct from that on a regular lattice, suggesting it belongs to a different universality class. We evaluate the exponent governing the bond fluctuations and our results agree with the Harris-Barghathi-Vojta criterium for relevant fluctuations.
\end{abstract} 

\maketitle

\section{Introduction}

In nonequilibrium systems, an absorbing-state phase transition occurs when a control parameter is varied, and the system experiences a phase transition from an active (fluctuating) state to an absorbing (with no fluctuations) state \cite{marrobook,henkel08}. Absorbing\sout{-state} phase transitions have attracted considerable interest in recent years, as they appear in a broad class of phenomena such as epidemic spreading \cite{pastor2014}, chemical reactions \cite{zgbqs}, population dynamics \cite{scp}, and others \cite{hinrichsen}.

As in equilibrium phase transitions, it is expected that critical phase transitions into absorbing states can be classified in a finite number of universality classes, depending on a few characteristics of the model, such as symmetries, conserved quantities, and dimensionality \cite{odor04}. 
Models such as the contact process (CP), with short-range interactions and without a conserved quantity
or symmetry beyond translational invariance, belong to the directed percolation (DP) universality class, the most prominent class of absorbing-state transitions \cite{gras,jans}.

The disorder is an unavoidable ingredient in real systems, so it is of primary interest to understand how spatially {\it quenched} (i.e., that does not change in time) disorder affects the critical behavior of absorbing-state phase transitions. 
On a regular lattice, the quenched disorder usually is introduced by dilution (random deletion
of sites or bonds)  \cite{noestPRB, adr-dic98, vojta06, DeOliveira2008} or by random spatial fluctuations  
of the control parameter \cite{durrett,salinas08,Amaral21}. In these cases, where the disorder is uncorrelated, rare regions that are locally super-critical (even when the whole system is sub-critical) emerge. The lifetime of these active rare regions grows exponentially with the domain size, leading to slow dynamics, with nonuniversal exponents, for some interval
of the control parameter $\lambda_c(0) < \lambda < \lambda_c$ with $\lambda_c(0)$ and
$\lambda_c$ being the critical points of the clean and disordered systems,
respectively. This kind of behavior marks a Griffiths phase (GP), and it appears in DP models with uncorrelated disorder, irrespective of the disorder strength \cite{DeOliveira2008,vojta09}. 

Strong disorder renormalization group methods and numerical studies showed that the activated-disorder behavior corresponds to the universality class of the random transverse Ising model  \cite{igloi1,igloi2}. 
These findings are in agreement with the heuristic Harris’ criterion \cite{harris74}, which states that uncorrelated quenched disorder is a relevant perturbation if $d\nu_\perp < 2$, where $d$ is the dimensionality, and $\nu_\perp$ is the correlation length exponent of the clean model. Here, it is noteworthy to mention that in DP class, this inequality is satisfied for all dimensions $d < 4$, since $\nu_\perp = 1.096854(4)$, $0.7333(75)$ and $0.584(5)$, for $d = 1, 2$ and
3, respectively \cite{henkel08}.

A further step is to understand the effects of {\it topological disorder}, which appears in nonperiodic, random structures. One example is the random lattice generated by the Voronoi-Delaunay (VD) triangulation \cite{okabe}, which is a bidimensional connected graph with a Poissonian distribution of connectivity with average degree $\overline q = 6$ (where we use the term degree to designate the node number of edges). In such a case, it was found that the disorder does not alter the critical behavior exhibited by the clean CP \cite{oliveira2,VD2}. These results are in clear contrast with the above mentioned results for uncorrelated disorder, which lead to an infinite-randomness critical point and strong GPs.

In order to determine the relevance of the disorder in these cases, extending early works by Luck \cite{luck93}, Barghathi and Vojta \cite{vojta14c} proposed what is known as the Harris-Barghathi-Vojta (HBV) criterion. It is based on the idea that the stability of the critical point is governed by the decay of spatial fluctuations of the local coordination numbers. Averaging the coordination numbers $q$ of the lattice nodes over coarse-grained blocks, the corresponding variance $\sigma_q$ decays algebraically with the block size, $\sigma_q(L_b)\sim L_b^{-a}$. Therefore, the HBV criterion states that quenched disorder is an irrelevant perturbation if the inequality $a\nu_\perp > 1$ holds (the exponent $a$ describes the decay of the fluctuations).
Note that, in the case of independent dilution, one has $a = d/2$, and the HBV criterion reduces to the original Harris criterion. Besides the HBV criterion has successfully explained the critical behavior in the VD triangulation, and in other examples with a topological disorder, very recently it was found examples of a class of systems that violate even the HBV criterion \cite{Schrauth2018,Schrauth2019,Schrauth2019b}.

In this work, we study the critical behavior of the CP in a distinct disordered structure, proposed by Hassan et al. \cite{Hassan2010},
the weighted planar stochastic (WPS) lattice. In comparison with the VD lattice, the WPS lattice
neither has a fixed cell size nor a fixed coordination number. On the other hand, the WPS lattice introduces topological disorder emerging from a multifractal structure \cite{Dayeen2016}, and its coordination number follows a power-law. Also, the WPS lattice resembles a crack tessellation \cite{crack}, which is applied in city modelling \cite{city}.

Hassan and Rahman \cite{Hassan2016} found that the isotropic percolation (bond and site) universality class on the WPS lattice is distinct from the one for all the known planar lattices \cite{Hsu}. It is suggested that the power-law distribution of the coordination number is the ingredient that changes the universality class. 
Liu et al. \cite{Liu} also found that an opinion dynamic model exhibits
distinct critical behavior on the WPS lattice, in contrast with 
the critical exponents of VD, which are the same
as two-dimensional Ising model \cite{Lima2000}. 

So, due to the importance of the DP universality class in the context of nonequilibrium phase transitions, and to give a further step in the understanding of the effects of distinct kinds of topological disorder in critical phenomena, in the present contribution, we aim to investigate whether disorder in the form of a coordination disorder exhibited by the WPS lattice alters the critical behavior of the contact process.

The remainder of this paper is organized as follows. In Sec. II, we present the model and methods employed in the work.
In Sec. III we show and discuss our simulation results; Sec. IV is
to summarize our conclusions.

\section{Models and methods}

\begin{figure*}
		\subfigure[]{\includegraphics[width=0.4\hsize]{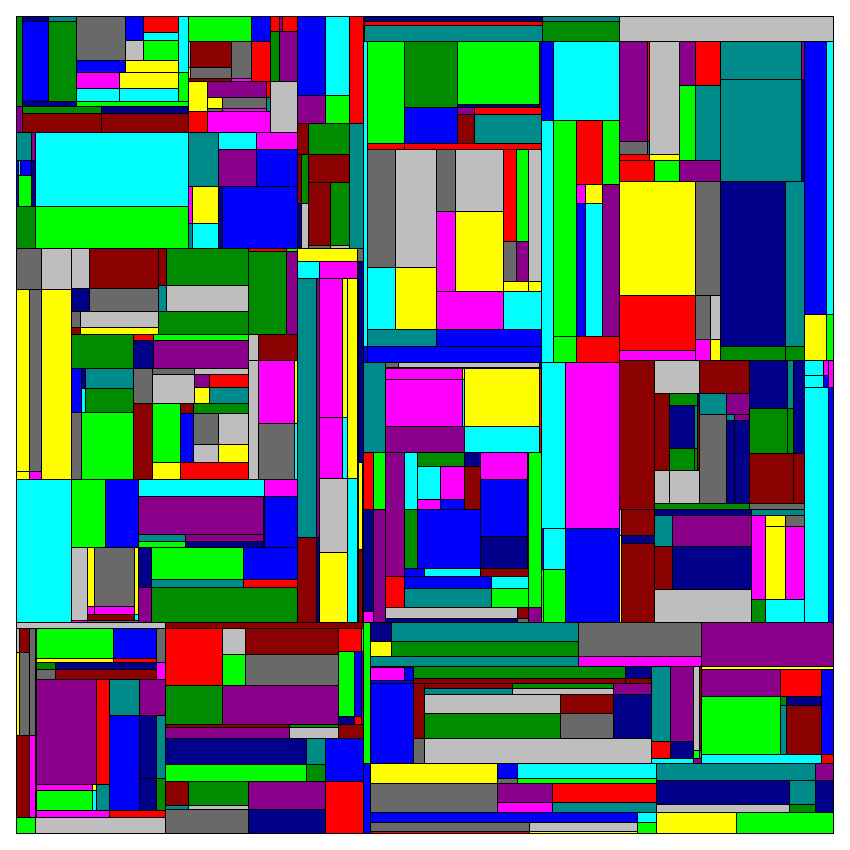}~~~}
		\subfigure[]{~~~\includegraphics[width=0.4\hsize]{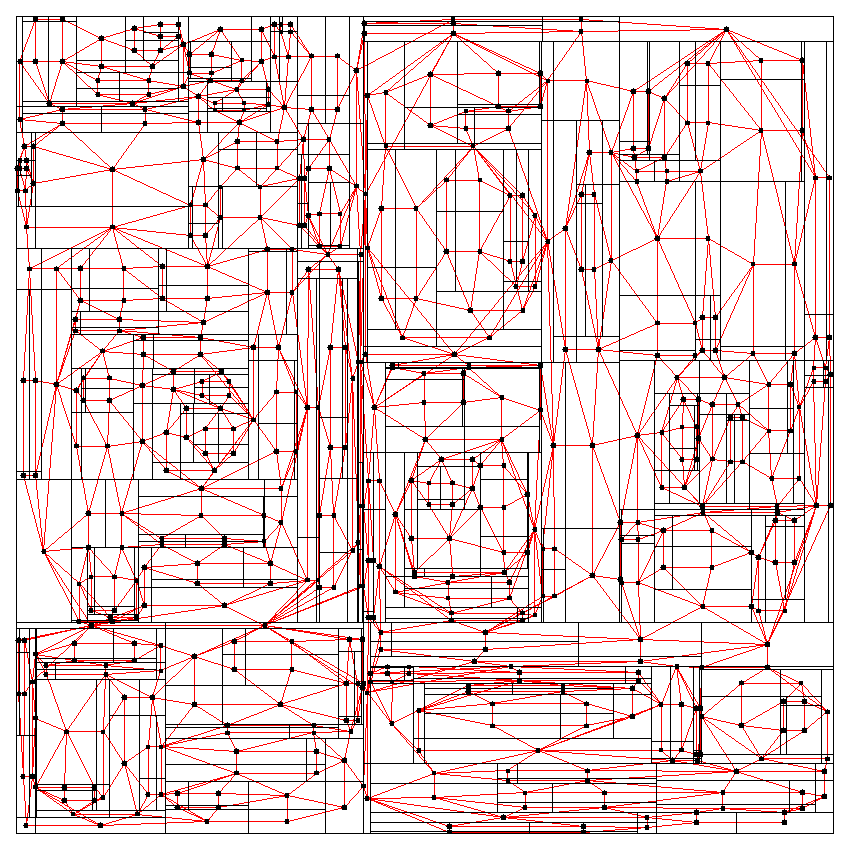}}
		\caption{\label{wpsl1}(Color online) (a) Snapshot of a block diagram obtained after $N_{it}=200$ steps of the WPSL construction. (b) The dual network of the block diagram  in (a). The dual network shown in (b) was obtained using open boundary condition.}
	\end{figure*}

The weighted planar stochastic lattice is built following the recipe in Ref.~\cite{Hassan2010}.  The planar lattice construction begins by taking a square and dividing it randomly into four smaller blocks. Then, these blocks are labeled by their respective areas in a clockwise fashion, starting from the upper left block. A block is chosen with a probability proportional to its area in the next steps. This step is repeated considering the block area weighting the probability, until the desired total iteration number (the process conserves the area of the starting square during the fragmentation process). After a number $N_{it}$ of iterations following such process, the resulting planar lattice has $3N_{it}+1$ blocks, and its dual lattice contains $N = 3N_{it}+1$ nodes. 
Each block is considered as a node and two blocks are connected if they share any part of its borders.
The total number of blocks sharing any part with a given block defines its neighbors number or degree.
We show a snapshot of a block diagram obtained after $N_{it}=200$ steps of the WPS construction in Fig.~\ref{wpsl1}(a), and in (b), its dual network.

\begin{figure}[h]
	{\includegraphics[width=0.95\hsize]{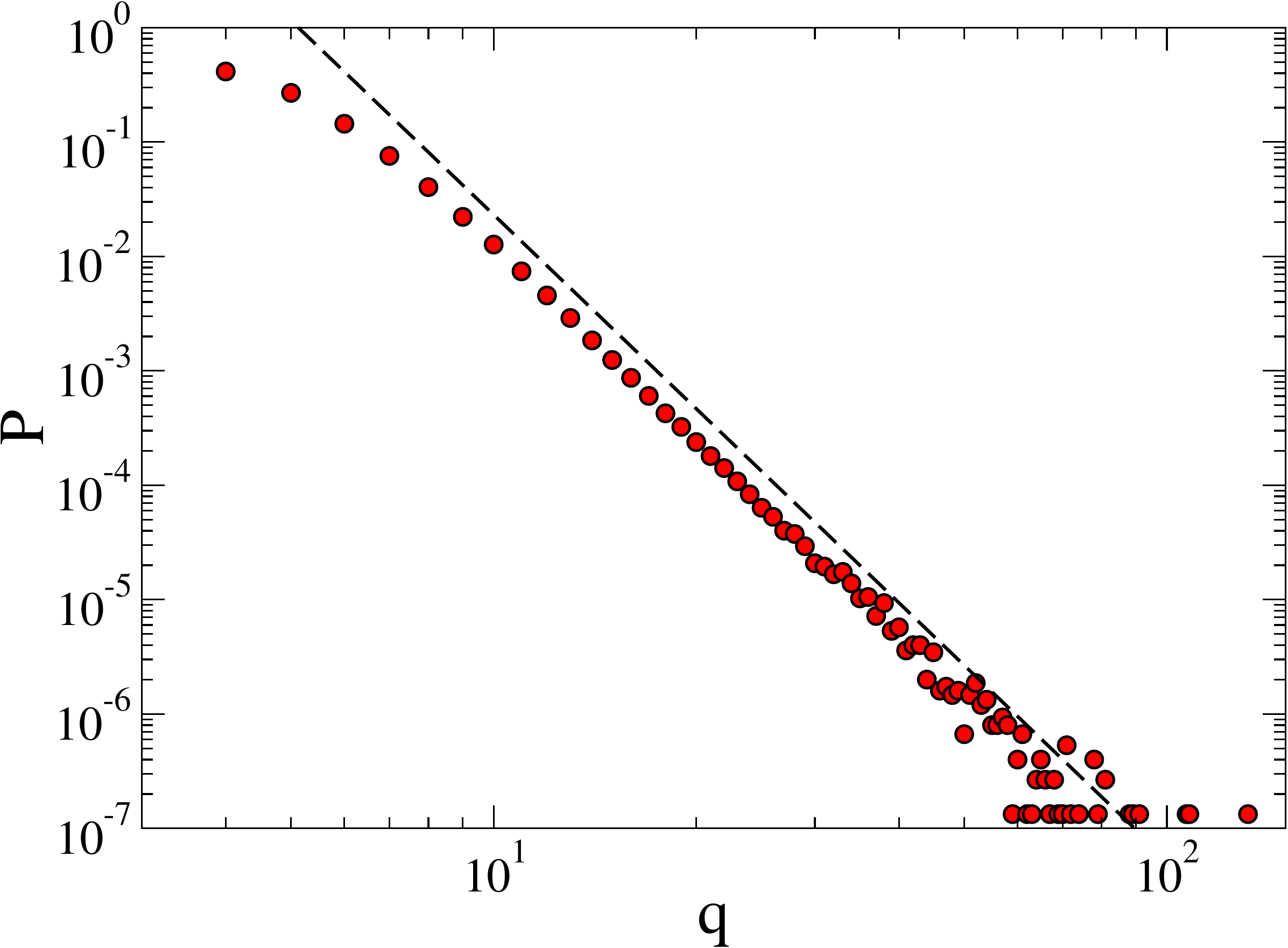}}
	\caption{\label{wpsl2}(Color online)  The degree probability distribution function. The dashed line is a power law with exponent $-5.635$. A mean over 100 networks with $N_{it}=10^4$ were used.}
\end{figure}

In comparison with the VD lattice, the WPS lattice neither has a fixed cell size nor a fixed degree. However, while in the VD lattice, the degree probability distribution function (PDF)  is peaked around its mean, in the dual of a WPS lattice, the degree PDF follows a power-law, $P(q)\sim q^{-5.6}$, as shown in Fig.~\ref{wpsl2}. The WPS lattice also presents size disorder as the size of its blocks exhibits a multifractal structure \cite{Dayeen2016}. In addition, the partitioning process can be used as a way to grow networks. So, we note the dual WPS network grows by a partitioning process, where four new nodes replace one old node at each iteration.
In the present work, all the simulations were performed using the dual WPS network (see Fig.~\ref{wpsl1}(b)) as the substrate for the standard CP.

The CP is a stochastic interacting particle system defined on a lattice, with each site $i$ either can be occupied ($\sigma_i(t) = 1$), or
vacant ($\sigma_i(t) = 0$). Particles are created at a site $i$ if at least one of its neighbors
is occupied: the transition from $\sigma_i = 0$ to $\sigma_i = 1$ occurs at rate $\lambda r$, where $r$ is the fraction
of nearest neighbors of site $i$ that are occupied. Particle annihilation, i.e, transitions from $\sigma_i = 1$ to $\sigma_i = 0$ occur at a unity rate, independent of
the neighboring sites. Therefore, the state $\sigma_i = 0$ for all $i$ is absorbing. For a critical value of the control parameter $\lambda=\lambda_c$, one observes a continuous phase transition between an active phase (with a positive density $\rho$ of occupied sites) to the absorbing state (with $\rho=0$).

In the computational scheme, we employ the usual simulation procedure: An occupied
site is chosen at random. Then, we choose between {\it annihilation} events
with probability $1/(1 + \lambda)$ and {\it creation} events with probability $\lambda/(1 + \lambda)$. In the case of annihilation, the chosen site is vacated, while, for creation events, one of its $q$ nearest-neighbor sites is selected at random and, if it is currently vacant, the chosen site becomes occupied. The
time increment associated with each such event is $\Delta t = 1/N_{occ}$, where $N_{occ}$ is the number of occupied sites just before the attempted transition.

To find the critical creation rate $\lambda_c$ we carry out spreading
analysis using WPS lattices with $N_it=10^6$ iterations starting the simulation with one node occupied.
After, we take averages restricted to the samples that did not visit an absorbing state. This was
performed employing the quasi-stationary simulation method \cite{qssimPRE,bjp}, which is based
on maintaining, and gradually updating, a set of configurations visited during the evolution.
The size of the set is $2000$ different configurations and this set is updated with
a probability $p_{rep}=0.1$.
If a transition to the absorbing state is imminent, the system is instead placed in one of the
saved configurations. Otherwise, the evolution is the same as that of a conventional simulation.
Finally, we perform initial decay investigations on large systems, starting with the fully occupied lattice.
In order to minimize the finite size effects, we have used periodic boundaries conditions
in the QS simulations and initial decay analysis. 

\section{Results and discussion}

\begin{figure}
	\includegraphics[width=0.95\hsize]{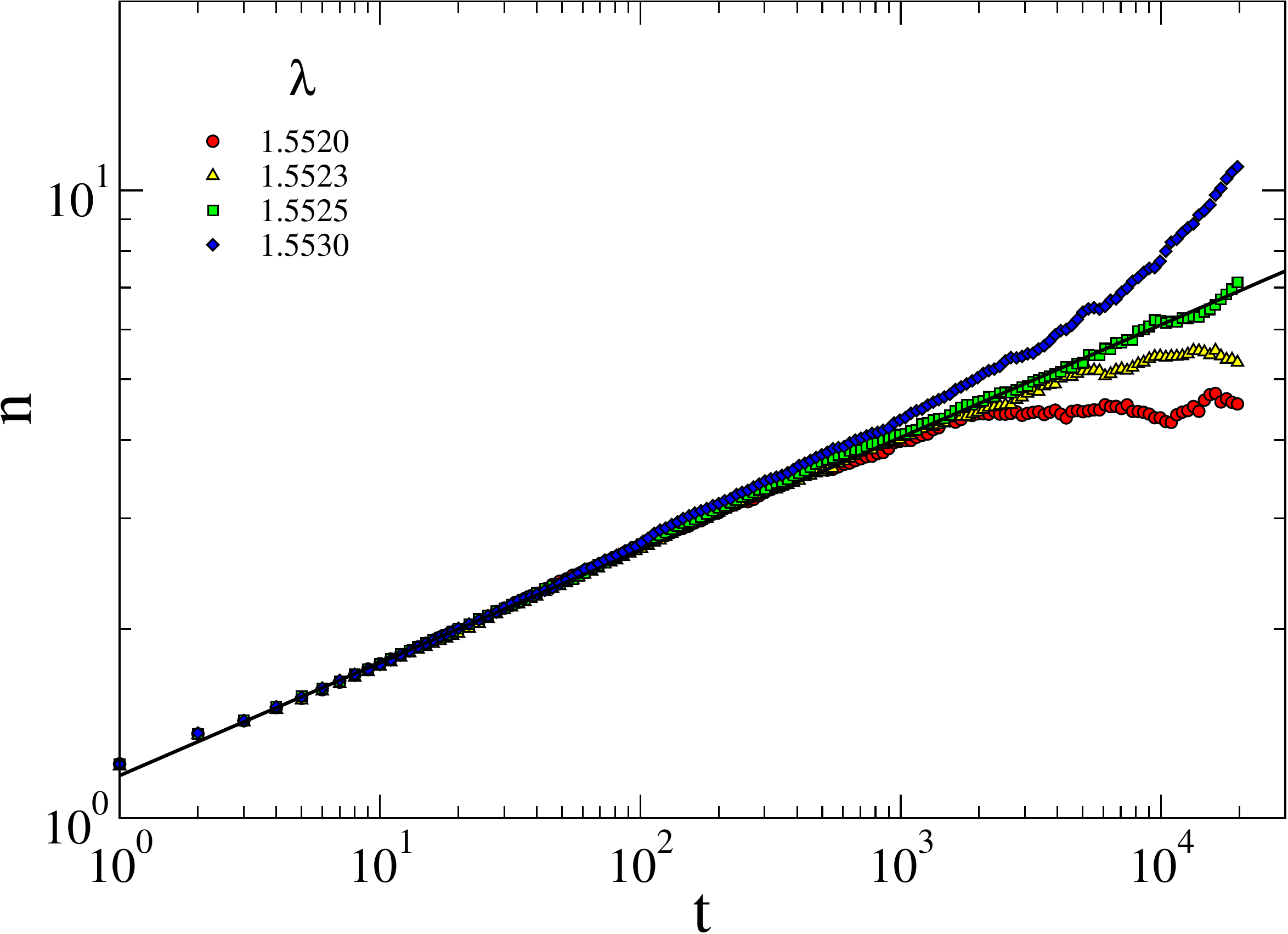}
	\caption{\label{n_t}(Color online)  Spreading of activity from a single seed. The straight line is a power law function with exponent 0.18.}
\end{figure}

The first step in analyzing our results is to determine the critical creation rate $\lambda_c$.
For this purpose, we study the number of active particles, $n(t)$ via spreading analysis using WPS
lattices with $10^6$ iterations and a mean over $5\times 10^3$ networks with $10$ run in each one.
The critical value $\lambda_c$ is then defined as the smallest $\lambda$ supporting asymptotic growth.
In Fig.~\ref{n_t}, we show the time evolution of the number of particles $n(t)$. From the data in the
figure, we obtain $\lambda_c=1.5525(5)$. We also observe the asymptotic evolution follows a power law,
\begin{equation}
n(t) \propto t^\eta,
\end{equation}
with $\eta=0.185(5)$. This value is distinct from the value $\eta= 0.2295(10)$ for the DP class.

Now, we perform extensive simulations of the CP on WPS random lattices with iteration numbers
$N_{it} = 128, 256, ..., 8192$, using the QS simulation method. Each realization of the process is
initialized with all sites occupied, and runs for at least $10^6$
time steps. Averages are taken in
the QS regime, after discarding an initial transient that depends on the system size. This
procedure is repeated for each realization of disorder (For each size studied, we performed
averages over 100 different lattices).

\begin{figure}
	\includegraphics[width=0.95\hsize]{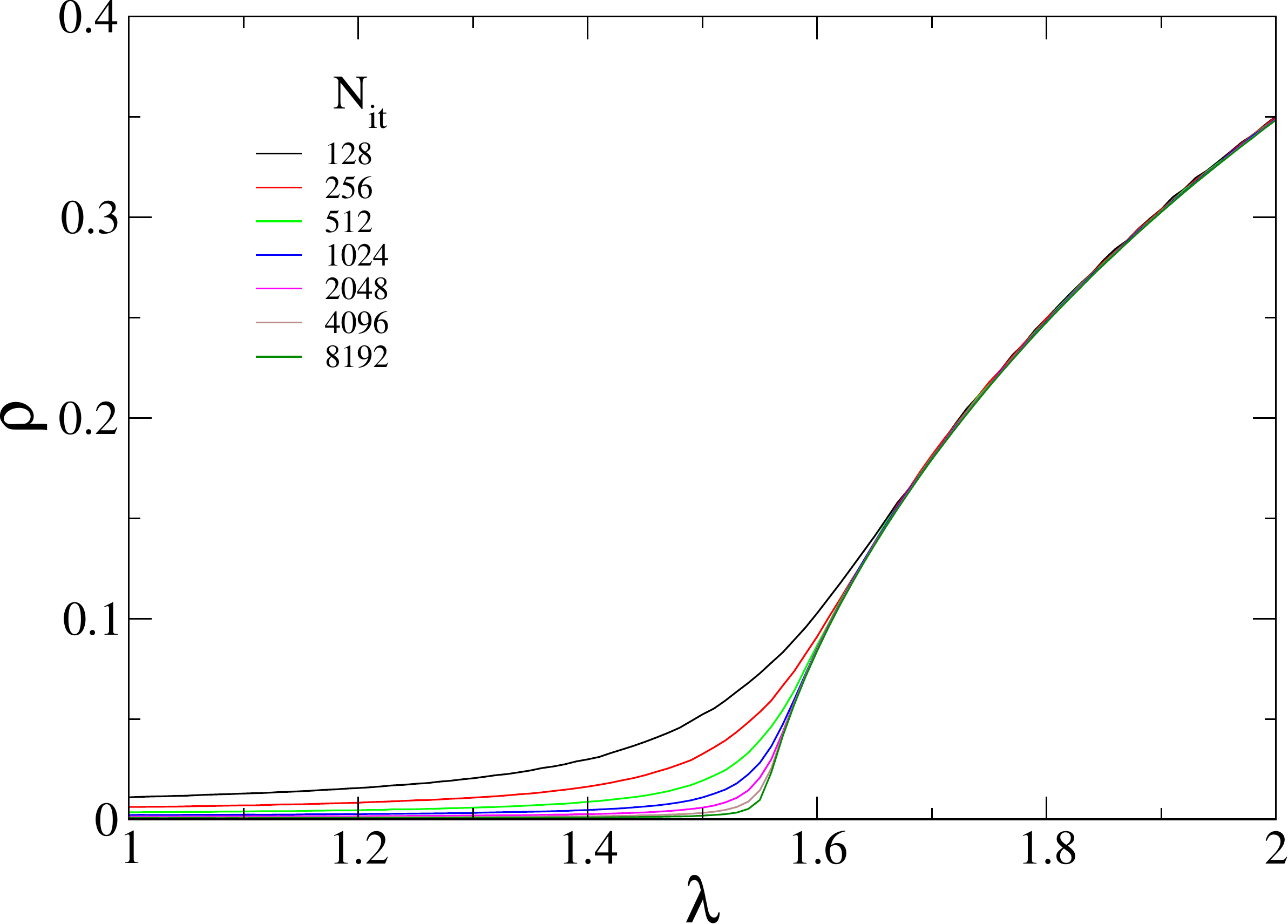}
	\caption{\label{rho}(Color online)  Quasistationary density as a function of the control parameter for different iterations number used (increasing from top to bottom).}
\end{figure}

In Fig.~\ref{rho} we show the quasistationary density $\rho$ as a function of the control parameter $\lambda$ for several values of $N_{it}$. We see, as expected, a continuous phase transition from an active to an absorbing state at $\lambda_c = 1.5525(5)$. Note that the threshold is slightly higher than those obtained for regular triangular lattices $\lambda_c=1.54780(5)$ and for the Voronoi triangulation $\lambda_c=1.54266(4)$, which also has $\overline{q}=6$. 

\begin{figure}
	\includegraphics[width=0.95\hsize]{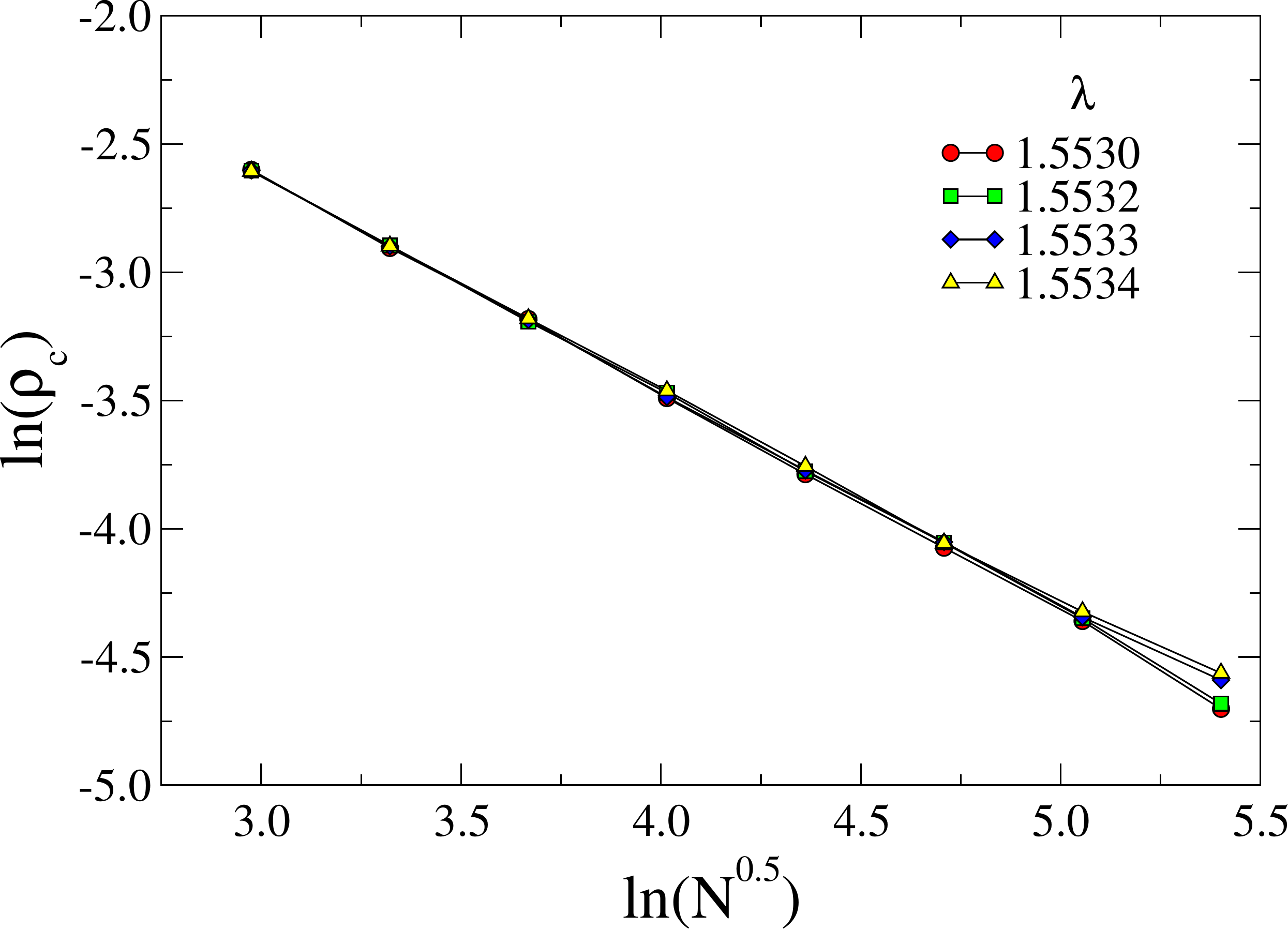}
	\caption{\label{rhoNit}(Color online)  Quasistationary density as a function of the number of nodes. Four values of the control parameter close to the critical value were used.}
\end{figure}

Analyzing the results shown in Fig.~\ref{rhoNit}, we see that at criticality, the quasistationary density of active sites, $\rho$ follows a power-law
\begin{equation}
\rho\propto L^{-\beta/\nu_\perp},
\end{equation}
where $L\propto \sqrt{N}$, is the linear system size. (Here we employed system of sizes generated after up to $N_{it} = 2^{16}$ interactions). 
Also, $\beta$ in the above equation is the critical exponent associated with the scaling of the density of active sites, $\rho\sim|\lambda-\lambda_c|^\beta$. And, finally, $\nu_\perp$ is the exponent associated with the divergence of the spatial correlation length $\varepsilon_\perp$ at criticality, $\varepsilon_\perp\sim |\lambda-\lambda_c|^{\nu_\perp}$.
From the data in Fig.~\ref{rhoNit}, we obtain the exponent ratio $\beta/\nu_\perp=0.84(1)$, which is distinct from the value 0.797(3) exhibited by models in the DP universality class.

\begin{figure}
	\includegraphics[width=0.95\hsize]{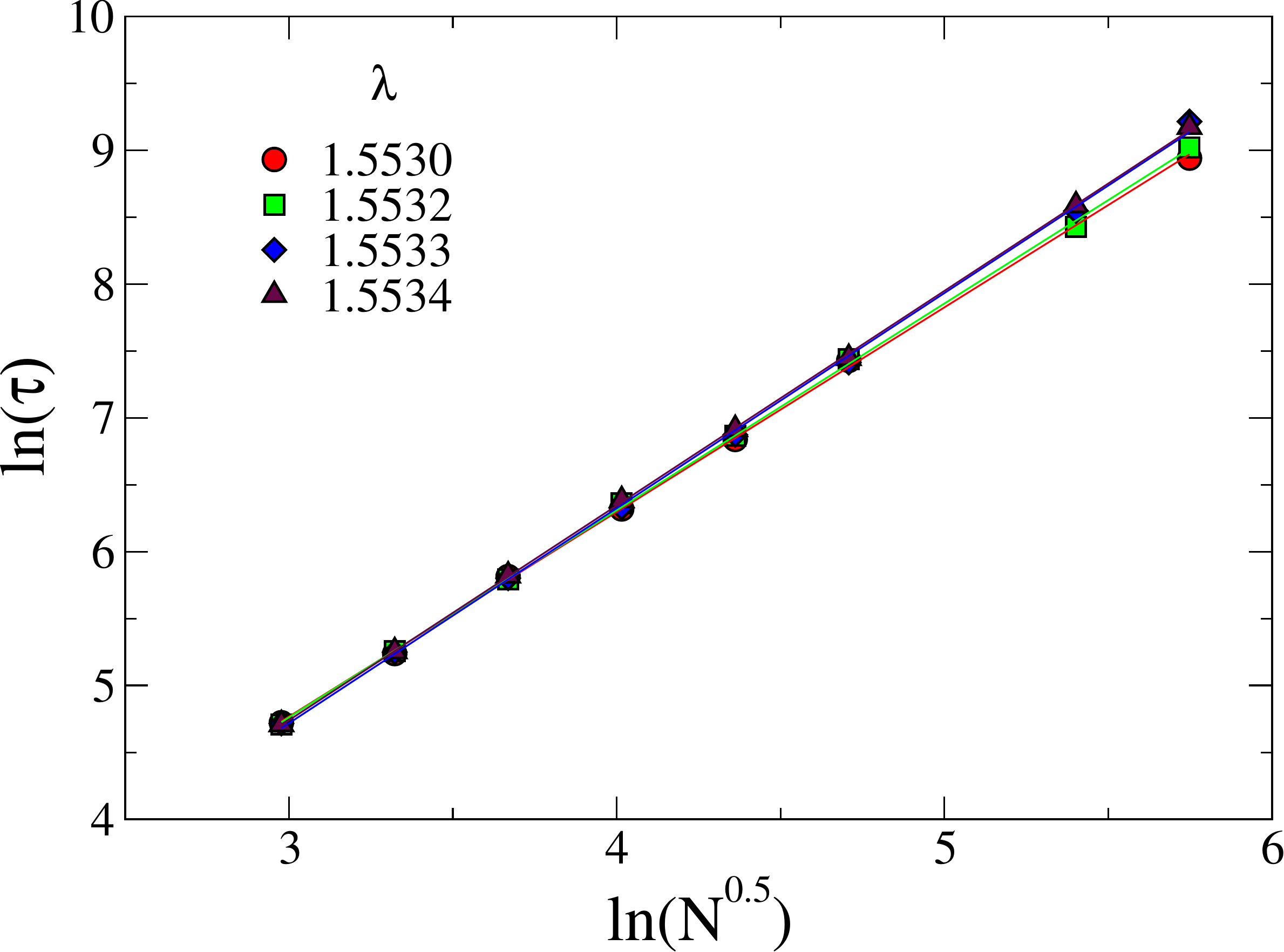}
	\caption{\label{tauL}(Color online)  Lifetime $\tau$ of the QS state as a function of the number of nodes. As in figure \ref{rhoNit} four values of $\lambda$ were used.}
\end{figure}

Now we turn to the dynamical exponent $z=\nu_\parallel/\nu_\perp$. We see the
lifetime of the QS state, $\tau$ (which we take as the mean time between two attempts to visit the
absorbing state in the QS simulation), follows
\begin{equation}
\tau\propto L^z
\end{equation}
at criticality. 
From the data in Fig.~\ref{tauL}, we obtain the exponent ratio $z=1.59(1)$, which is distinct from the value 1.7674(6) exhibited by models in the DP universality class.

\begin{figure}
	\includegraphics[width=0.95\hsize]{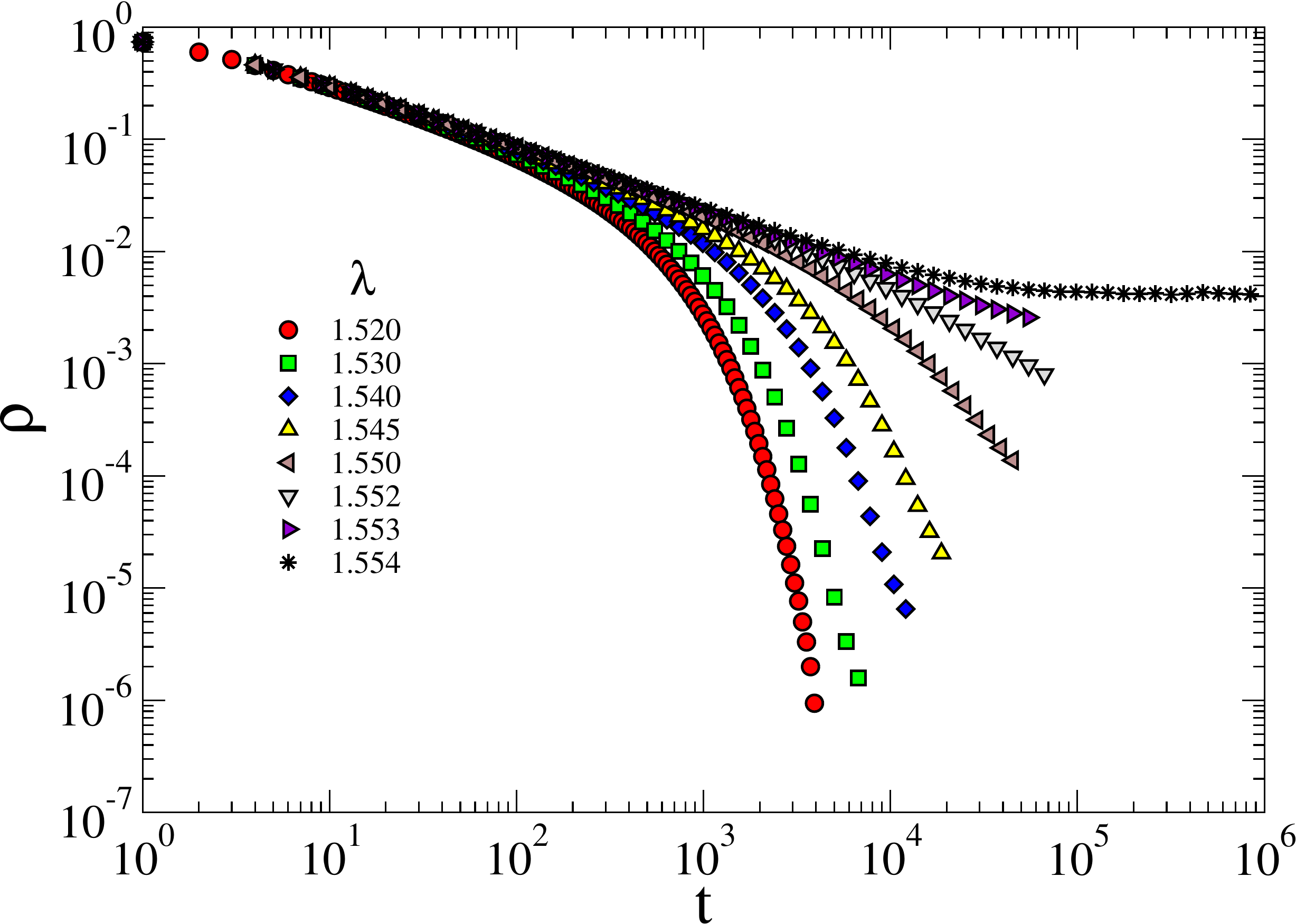}
	\caption{\label{decai}(Color online) Initial decay simulations: particle density $\rho$ vs. $t$.
(System size $N = 3 \times 10^6$ ) .}
\end{figure}

We also search for a Griffiths phase, which would be a remark of activated dynamics. A Griffiths phase is a region inside the subcritical phase where the long-time decay of $\rho$ towards the extinction is algebraic (with non-universal exponents), rather than exponential, that emerges in the presence of relevant disorder in the activated dynamics scenario \cite{vojta06}.
So, in Fig.~\ref{decai}, we present results from initial decay simulations, where the system
starts its dynamics from a fully occupied lattice. We only observe exponential decay in the subcritical phase, without any sign of a Griffiths phase. At criticality, we note
\begin{equation}
\rho\propto t^{-\delta},
\end{equation}
and we obtain $\delta=0.57(3)$, which is far from the DP value,  $\delta=0.4505(1)$.

In summary, our results reveal that the absorbing phase transition of the contact process defined on
the WPS lattice does not belong to the directed percolation universality
class. It is important to mention that, while the exponents differ from those
obtained for the DP class, they still obey the scaling relation $4\delta+2\eta=2d/z$ \cite{marrobook}.
We also did not find any hallmark of a Griffiths phase, however, we can not rule out its existence if
larger systems sizes are considered. 

\begin{figure}
	{\includegraphics[width=0.95\hsize]{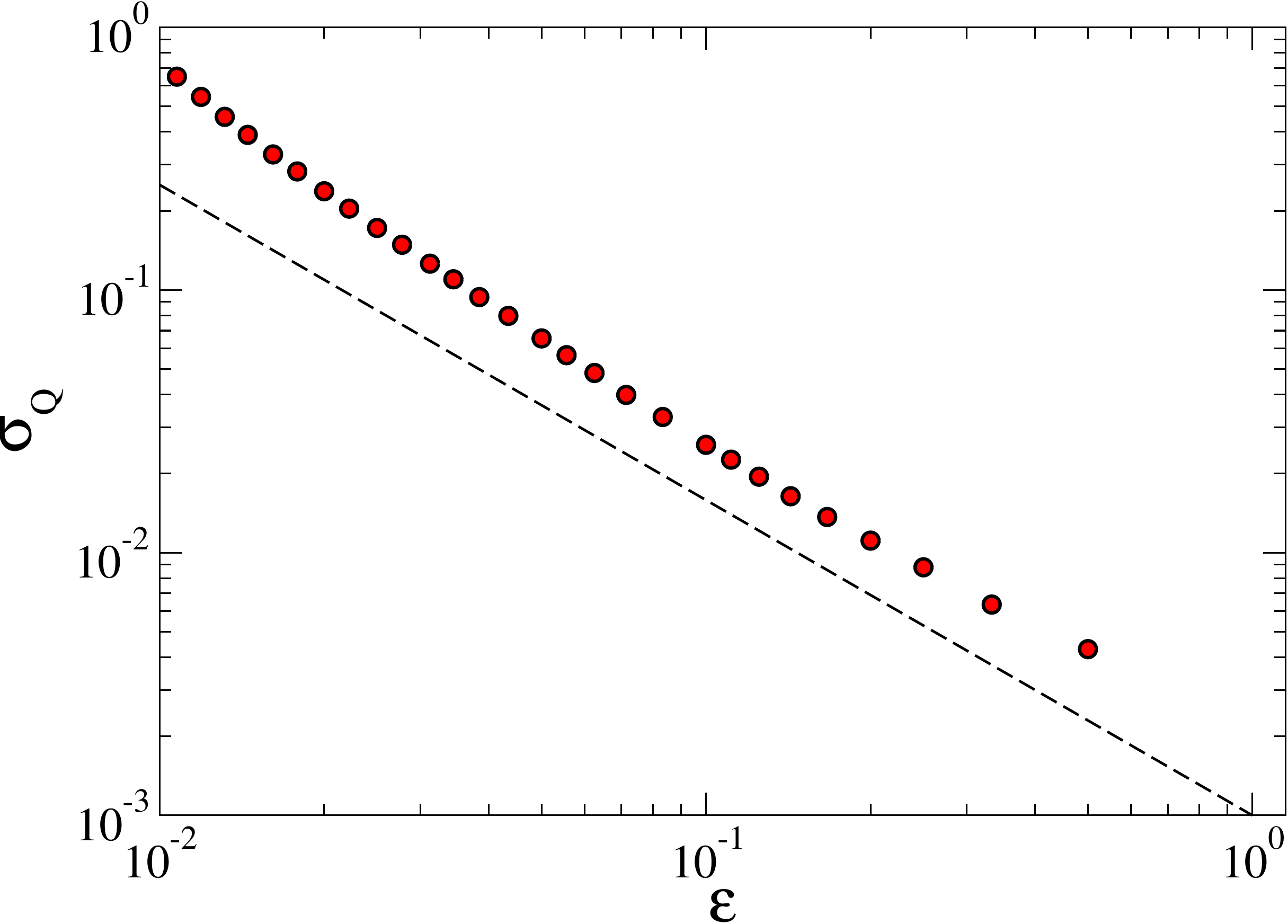}}
	\caption{\label{sigma}(Color online)  	Fluctuation of the mean degree (or connectivity) as a function
		of the grid size, bottom panel. The dashed line is a power law decay with exponent $1.2$. A mean
		over 100 network with $N_{it}=10^4$ were used.}
\end{figure}
Now, to check the agreement of our simulation results with the predictions given by the HBV criterion,
we evaluate the wandering exponent of the WPS lattice. We investigate the degree fluctuation, the system
is divided in blocks of size $\epsilon$ and we calculate the block-averaged degree and its standard
deviation $\sigma_Q$ \cite{vojta14c}. In Fig.~\ref{sigma} we plot the fluctuation of the mean connection
$\sigma_Q$ as a function of the grid size $\epsilon$. We obtain the exponent $a=1.2$, that yields
$a\nu_\perp=0.88$ for $d=2$. Therefore, according to the HBV criterion, the disorder presented by
the WPS lattice should be {\it relevant}, which is in agreement with our findings.  

\section{Conclusions}

We performed extensive large-scale simulations of the contact process on the Weighted Planar Stochastic Lattice, which presents multifractality and coordination number power-law distributed. We evaluated several critical exponents and found they are distinct from those of the DP class. So, our results reveal that the topological disorder exhibited by the WPS lattice is a relevant perturbation to the DP universality class. By numerically evaluating the exponent governing the fluctuations due to the lattice topology, we notice our findings agree with the predictions of the HBV criterion.

As a final remark, despite the quenched disorder present in the WPS dual network, it is important to mention that for infinite dimensional power-law networks is known that the critical behavior depends on the degree distribution exponent \cite{silvioQS}. Therefore, it would be interesting to investigate if some kind of modification in the algorithm of the partitioning process can generate dual networks with distinct exponents. This would be useful in clarifying the role of long-range connections and the quenched disorder in changing the critical behavior of the CP on WPS networks.

\section{acknowledgement}
We acknowledge the anonymous referees for several useful suggestions.
This work was supported by CNPq and FAPEMIG, Brazilian agencies.

\bibliography{cp_on_wpsl}

\begin{thebibliography}{39}%
\makeatletter
\providecommand \@ifxundefined [1]{%
 \@ifx{#1\undefined}
}%
\providecommand \@ifnum [1]{%
 \ifnum #1\expandafter \@firstoftwo
 \else \expandafter \@secondoftwo
 \fi
}%
\providecommand \@ifx [1]{%
 \ifx #1\expandafter \@firstoftwo
 \else \expandafter \@secondoftwo
 \fi
}%
\providecommand \natexlab [1]{#1}%
\providecommand \enquote  [1]{``#1''}%
\providecommand \bibnamefont  [1]{#1}%
\providecommand \bibfnamefont [1]{#1}%
\providecommand \citenamefont [1]{#1}%
\providecommand \href@noop [0]{\@secondoftwo}%
\providecommand \href [0]{\begingroup \@sanitize@url \@href}%
\providecommand \@href[1]{\@@startlink{#1}\@@href}%
\providecommand \@@href[1]{\endgroup#1\@@endlink}%
\providecommand \@sanitize@url [0]{\catcode `\\12\catcode `\$12\catcode
  `\&12\catcode `\#12\catcode `\^12\catcode `\_12\catcode `\%12\relax}%
\providecommand \@@startlink[1]{}%
\providecommand \@@endlink[0]{}%
\providecommand \url  [0]{\begingroup\@sanitize@url \@url }%
\providecommand \@url [1]{\endgroup\@href {#1}{\urlprefix }}%
\providecommand \urlprefix  [0]{URL }%
\providecommand \Eprint [0]{\href }%
\providecommand \doibase [0]{http://dx.doi.org/}%
\providecommand \selectlanguage [0]{\@gobble}%
\providecommand \bibinfo  [0]{\@secondoftwo}%
\providecommand \bibfield  [0]{\@secondoftwo}%
\providecommand \translation [1]{[#1]}%
\providecommand \BibitemOpen [0]{}%
\providecommand \bibitemStop [0]{}%
\providecommand \bibitemNoStop [0]{.\EOS\space}%
\providecommand \EOS [0]{\spacefactor3000\relax}%
\providecommand \BibitemShut  [1]{\csname bibitem#1\endcsname}%
\let\auto@bib@innerbib\@empty
\bibitem [{\citenamefont {Marro}\ and\ \citenamefont
  {Dickman}(1999)}]{marrobook}%
  \BibitemOpen
  \bibfield  {author} {\bibinfo {author} {\bibfnamefont {J.}~\bibnamefont
  {Marro}}\ and\ \bibinfo {author} {\bibfnamefont {R.}~\bibnamefont
  {Dickman}},\ }\href@noop {} {\emph {\bibinfo {title} {Nonequilibrium phase
  transitions in lattice models}}}\ (\bibinfo  {publisher} {Cambridge
  University Press},\ \bibinfo {address} {Cambridge},\ \bibinfo {year}
  {1999})\BibitemShut {NoStop}%
\bibitem [{\citenamefont {Henkel}\ \emph {et~al.}(2008)\citenamefont {Henkel},
  \citenamefont {Hinrichsen}, \citenamefont {L{\"u}},\ and\ \citenamefont
  {Pleimling}}]{henkel08}%
  \BibitemOpen
  \bibfield  {author} {\bibinfo {author} {\bibfnamefont {M.}~\bibnamefont
  {Henkel}}, \bibinfo {author} {\bibfnamefont {H.}~\bibnamefont {Hinrichsen}},
  \bibinfo {author} {\bibfnamefont {S.}~\bibnamefont {L{\"u}}}, \ and\ \bibinfo
  {author} {\bibfnamefont {M.}~\bibnamefont {Pleimling}},\ }\href@noop {}
  {\emph {\bibinfo {title} {Non-equilibrium phase transitions}}},\
  Vol.~\bibinfo {volume} {1}\ (\bibinfo  {publisher} {Springer},\ \bibinfo
  {address} {Dordrecht, Netherlands},\ \bibinfo {year} {2008})\BibitemShut
  {NoStop}%
\bibitem [{\citenamefont {Pastor-Satorras}\ \emph {et~al.}(2015)\citenamefont
  {Pastor-Satorras}, \citenamefont {Castellano}, \citenamefont {Van~Mieghem},\
  and\ \citenamefont {Vespignani}}]{pastor2014}%
  \BibitemOpen
  \bibfield  {author} {\bibinfo {author} {\bibfnamefont {R.}~\bibnamefont
  {Pastor-Satorras}}, \bibinfo {author} {\bibfnamefont {C.}~\bibnamefont
  {Castellano}}, \bibinfo {author} {\bibfnamefont {P.}~\bibnamefont
  {Van~Mieghem}}, \ and\ \bibinfo {author} {\bibfnamefont {A.}~\bibnamefont
  {Vespignani}},\ }\href {\doibase 10.1103/RevModPhys.87.925} {\bibfield
  {journal} {\bibinfo  {journal} {Rev. Mod. Phys.}\ }\textbf {\bibinfo {volume}
  {87}},\ \bibinfo {pages} {925} (\bibinfo {year} {2015})}\BibitemShut
  {NoStop}%
\bibitem [{\citenamefont {{de Oliveira}}\ and\ \citenamefont
  {Dickman}(2004)}]{zgbqs}%
  \BibitemOpen
  \bibfield  {author} {\bibinfo {author} {\bibfnamefont {M.~M.}\ \bibnamefont
  {{de Oliveira}}}\ and\ \bibinfo {author} {\bibfnamefont {R.}~\bibnamefont
  {Dickman}},\ }\href {\doibase 10.1016/j.physa.2004.06.155} {\bibfield
  {journal} {\bibinfo  {journal} {Physica A: Statistical Mechanics and its
  Applications}\ }\textbf {\bibinfo {volume} {343}} (\bibinfo {year} {2004}),\
  10.1016/j.physa.2004.06.155}\BibitemShut {NoStop}%
\bibitem [{\citenamefont {{de Oliveira}}\ \emph {et~al.}(2012)\citenamefont
  {{de Oliveira}}, \citenamefont {Dos~Santos},\ and\ \citenamefont
  {Dickman}}]{scp}%
  \BibitemOpen
  \bibfield  {author} {\bibinfo {author} {\bibfnamefont {M.~M.}\ \bibnamefont
  {{de Oliveira}}}, \bibinfo {author} {\bibfnamefont {R.~V.}\ \bibnamefont
  {Dos~Santos}}, \ and\ \bibinfo {author} {\bibfnamefont {R.}~\bibnamefont
  {Dickman}},\ }\href@noop {} {\bibfield  {journal} {\bibinfo  {journal} {Phys.
  Rev. E}\ }\textbf {\bibinfo {volume} {86}},\ \bibinfo {pages} {011121}
  (\bibinfo {year} {2012})}\BibitemShut {NoStop}%
\bibitem [{\citenamefont {Hinrichsen}(2000)}]{hinrichsen}%
  \BibitemOpen
  \bibfield  {author} {\bibinfo {author} {\bibfnamefont {H.}~\bibnamefont
  {Hinrichsen}},\ }\href@noop {} {\bibfield  {journal} {\bibinfo  {journal}
  {Advances in Physics}\ }\textbf {\bibinfo {volume} {49}},\ \bibinfo {pages}
  {815} (\bibinfo {year} {2000})}\BibitemShut {NoStop}%
\bibitem [{\citenamefont {\'Odor}(2004)}]{odor04}%
  \BibitemOpen
  \bibfield  {author} {\bibinfo {author} {\bibfnamefont {G.}~\bibnamefont
  {\'Odor}},\ }\href@noop {} {\bibfield  {journal} {\bibinfo  {journal} {Rev.
  Mod. Phys.}\ }\textbf {\bibinfo {volume} {76}},\ \bibinfo {pages} {663}
  (\bibinfo {year} {2004})}\BibitemShut {NoStop}%
\bibitem [{\citenamefont {Janssen}(1981)}]{gras}%
  \BibitemOpen
  \bibfield  {author} {\bibinfo {author} {\bibfnamefont {H.}~\bibnamefont
  {Janssen}},\ }\href@noop {} {\bibfield  {journal} {\bibinfo  {journal} {Z
  Phys. B}\ }\textbf {\bibinfo {volume} {42}},\ \bibinfo {pages} {151}
  (\bibinfo {year} {1981})}\BibitemShut {NoStop}%
\bibitem [{\citenamefont {Grassberger}(1982)}]{jans}%
  \BibitemOpen
  \bibfield  {author} {\bibinfo {author} {\bibfnamefont {P.}~\bibnamefont
  {Grassberger}},\ }\href@noop {} {\bibfield  {journal} {\bibinfo  {journal} {Z
  Phys. B}\ }\textbf {\bibinfo {volume} {47}},\ \bibinfo {pages} {365}
  (\bibinfo {year} {1982})}\BibitemShut {NoStop}%
\bibitem [{\citenamefont {Noest}(1988)}]{noestPRB}%
  \BibitemOpen
  \bibfield  {author} {\bibinfo {author} {\bibfnamefont {A.~J.}\ \bibnamefont
  {Noest}},\ }\href {\doibase 10.1103/PhysRevB.38.2715} {\bibfield  {journal}
  {\bibinfo  {journal} {Phys. Rev. B}\ }\textbf {\bibinfo {volume} {38}},\
  \bibinfo {pages} {2715} (\bibinfo {year} {1988})}\BibitemShut {NoStop}%
\bibitem [{\citenamefont {Dickman}\ and\ \citenamefont
  {Moreira}(1998)}]{adr-dic98}%
  \BibitemOpen
  \bibfield  {author} {\bibinfo {author} {\bibfnamefont {R.}~\bibnamefont
  {Dickman}}\ and\ \bibinfo {author} {\bibfnamefont {A.~G.}\ \bibnamefont
  {Moreira}},\ }\href {\doibase 10.1103/PhysRevE.57.1263} {\bibfield  {journal}
  {\bibinfo  {journal} {Phys. Rev. E}\ }\textbf {\bibinfo {volume} {57}},\
  \bibinfo {pages} {1263} (\bibinfo {year} {1998})}\BibitemShut {NoStop}%
\bibitem [{\citenamefont {Vojta}\ and\ \citenamefont {Lee}(2006)}]{vojta06}%
  \BibitemOpen
  \bibfield  {author} {\bibinfo {author} {\bibfnamefont {T.}~\bibnamefont
  {Vojta}}\ and\ \bibinfo {author} {\bibfnamefont {M.~Y.}\ \bibnamefont
  {Lee}},\ }\href {\doibase 10.1103/PhysRevLett.96.035701} {\bibfield
  {journal} {\bibinfo  {journal} {Phys. Rev. Lett.}\ }\textbf {\bibinfo
  {volume} {96}},\ \bibinfo {pages} {035701} (\bibinfo {year}
  {2006})}\BibitemShut {NoStop}%
\bibitem [{\citenamefont {{de Oliveira}}\ and\ \citenamefont
  {Ferreira}(2008)}]{DeOliveira2008}%
  \BibitemOpen
  \bibfield  {author} {\bibinfo {author} {\bibfnamefont {M.~M.}\ \bibnamefont
  {{de Oliveira}}}\ and\ \bibinfo {author} {\bibfnamefont {S.~C.}\ \bibnamefont
  {Ferreira}},\ }\href {\doibase 10.1088/1742-5468/2008/11/P11001} {\bibfield
  {journal} {\bibinfo  {journal} {J. Stat. Mech.: Theor. Exp.}\ ,\ \bibinfo
  {pages} {P11001}} (\bibinfo {year} {2008})},\ \Eprint
  {http://arxiv.org/abs/1101.1100} {1101.1100} \BibitemShut {NoStop}%
\bibitem [{\citenamefont {Bramson}\ \emph {et~al.}(1991)\citenamefont
  {Bramson}, \citenamefont {Durrett},\ and\ \citenamefont
  {Schonmann}}]{durrett}%
  \BibitemOpen
  \bibfield  {author} {\bibinfo {author} {\bibfnamefont {M.}~\bibnamefont
  {Bramson}}, \bibinfo {author} {\bibfnamefont {R.}~\bibnamefont {Durrett}}, \
  and\ \bibinfo {author} {\bibfnamefont {R.~H.}\ \bibnamefont {Schonmann}},\
  }\href@noop {} {\bibfield  {journal} {\bibinfo  {journal} {Ann. Probab.}\
  }\textbf {\bibinfo {volume} {19}},\ \bibinfo {pages} {pp. 960} (\bibinfo
  {year} {1991})}\BibitemShut {NoStop}%
\bibitem [{\citenamefont {Faria}\ \emph {et~al.}(2008)\citenamefont {Faria},
  \citenamefont {Ribeiro},\ and\ \citenamefont {S.~A.~Salinas}}]{salinas08}%
  \BibitemOpen
  \bibfield  {author} {\bibinfo {author} {\bibfnamefont {M.~S.}\ \bibnamefont
  {Faria}}, \bibinfo {author} {\bibfnamefont {D.~J.}\ \bibnamefont {Ribeiro}},
  \ and\ \bibinfo {author} {\bibfnamefont {J.}~\bibnamefont {S.~A.~Salinas}},\
  }\href@noop {} {\bibfield  {journal} {\bibinfo  {journal} {J. Stat. Mech.:
  Theor. Exp.}\ ,\ \bibinfo {pages} {P11001}} (\bibinfo {year}
  {2008})}\BibitemShut {NoStop}%
\bibitem [{\citenamefont {Amaral}\ and\ \citenamefont
  {de~Oliveira}(2021)}]{Amaral21}%
  \BibitemOpen
  \bibfield  {author} {\bibinfo {author} {\bibfnamefont {M.~A.}\ \bibnamefont
  {Amaral}}\ and\ \bibinfo {author} {\bibfnamefont {M.~M.}\ \bibnamefont
  {de~Oliveira}},\ }\href {\doibase 10.1103/PhysRevE.104.064102} {\bibfield
  {journal} {\bibinfo  {journal} {Phys. Rev. E}\ }\textbf {\bibinfo {volume}
  {104}},\ \bibinfo {pages} {064102} (\bibinfo {year} {2021})}\BibitemShut
  {NoStop}%
\bibitem [{\citenamefont {Vojta}\ \emph {et~al.}(2009)\citenamefont {Vojta},
  \citenamefont {Farquhar},\ and\ \citenamefont {Mast}}]{vojta09}%
  \BibitemOpen
  \bibfield  {author} {\bibinfo {author} {\bibfnamefont {T.}~\bibnamefont
  {Vojta}}, \bibinfo {author} {\bibfnamefont {A.}~\bibnamefont {Farquhar}}, \
  and\ \bibinfo {author} {\bibfnamefont {J.}~\bibnamefont {Mast}},\ }\href
  {\doibase 10.1103/PhysRevE.79.011111} {\bibfield  {journal} {\bibinfo
  {journal} {Phys. Rev. E}\ }\textbf {\bibinfo {volume} {79}},\ \bibinfo
  {pages} {011111} (\bibinfo {year} {2009})}\BibitemShut {NoStop}%
\bibitem [{\citenamefont {Hooyberghs}\ \emph {et~al.}(2003)\citenamefont
  {Hooyberghs}, \citenamefont {Igl\'oi},\ and\ \citenamefont
  {Vanderzande}}]{igloi1}%
  \BibitemOpen
  \bibfield  {author} {\bibinfo {author} {\bibfnamefont {J.}~\bibnamefont
  {Hooyberghs}}, \bibinfo {author} {\bibfnamefont {F.}~\bibnamefont {Igl\'oi}},
  \ and\ \bibinfo {author} {\bibfnamefont {C.}~\bibnamefont {Vanderzande}},\
  }\href {\doibase 10.1103/PhysRevLett.90.100601} {\bibfield  {journal}
  {\bibinfo  {journal} {Phys. Rev. Lett.}\ }\textbf {\bibinfo {volume} {90}},\
  \bibinfo {pages} {100601} (\bibinfo {year} {2003})}\BibitemShut {NoStop}%
\bibitem [{\citenamefont {Hooyberghs}\ \emph {et~al.}(2004)\citenamefont
  {Hooyberghs}, \citenamefont {Igl\'oi},\ and\ \citenamefont
  {Vanderzande}}]{igloi2}%
  \BibitemOpen
  \bibfield  {author} {\bibinfo {author} {\bibfnamefont {J.}~\bibnamefont
  {Hooyberghs}}, \bibinfo {author} {\bibfnamefont {F.}~\bibnamefont {Igl\'oi}},
  \ and\ \bibinfo {author} {\bibfnamefont {C.}~\bibnamefont {Vanderzande}},\
  }\href {\doibase 10.1103/PhysRevE.69.066140} {\bibfield  {journal} {\bibinfo
  {journal} {Phys. Rev. E}\ }\textbf {\bibinfo {volume} {69}},\ \bibinfo
  {pages} {066140} (\bibinfo {year} {2004})}\BibitemShut {NoStop}%
\bibitem [{\citenamefont {Harris}(1974)}]{harris74}%
  \BibitemOpen
  \bibfield  {author} {\bibinfo {author} {\bibfnamefont {A.~B.}\ \bibnamefont
  {Harris}},\ }\href@noop {} {\bibfield  {journal} {\bibinfo  {journal} {J.
  Phys. C: Solid State Physics}\ }\textbf {\bibinfo {volume} {7}},\ \bibinfo
  {pages} {1671} (\bibinfo {year} {1974})}\BibitemShut {NoStop}%
\bibitem [{\citenamefont {Okabe}\ \emph {et~al.}(2000)\citenamefont {Okabe},
  \citenamefont {Boots}, \citenamefont {Sugihara},\ and\ \citenamefont
  {Chiu}}]{okabe}%
  \BibitemOpen
  \bibfield  {author} {\bibinfo {author} {\bibfnamefont {A.}~\bibnamefont
  {Okabe}}, \bibinfo {author} {\bibfnamefont {B.}~\bibnamefont {Boots}},
  \bibinfo {author} {\bibfnamefont {K.}~\bibnamefont {Sugihara}}, \ and\
  \bibinfo {author} {\bibfnamefont {S.~N.}\ \bibnamefont {Chiu}},\ }\href@noop
  {} {\emph {\bibinfo {title} {Spatial tessellations: concepts and applications
  of Vo\-ro\-noi diagrams}}}\ (\bibinfo  {publisher} {John Wiley and Sons
  Lts.},\ \bibinfo {address} {Cichester},\ \bibinfo {year} {2000})\BibitemShut
  {NoStop}%
\bibitem [{\citenamefont {{de Oliveira}}\ \emph {et~al.}(2008)\citenamefont
  {{de Oliveira}}, \citenamefont {Alves}, \citenamefont {Ferreira},\ and\
  \citenamefont {Dickman}}]{oliveira2}%
  \BibitemOpen
  \bibfield  {author} {\bibinfo {author} {\bibfnamefont {M.~M.}\ \bibnamefont
  {{de Oliveira}}}, \bibinfo {author} {\bibfnamefont {S.~G.}\ \bibnamefont
  {Alves}}, \bibinfo {author} {\bibfnamefont {S.~C.}\ \bibnamefont {Ferreira}},
  \ and\ \bibinfo {author} {\bibfnamefont {R.}~\bibnamefont {Dickman}},\ }\href
  {\doibase 10.1103/PhysRevE.78.031133} {\bibfield  {journal} {\bibinfo
  {journal} {Phys. Rev. E}\ }\textbf {\bibinfo {volume} {78}},\ \bibinfo
  {pages} {031133} (\bibinfo {year} {2008})}\BibitemShut {NoStop}%
\bibitem [{\citenamefont {{De Oliveira}}\ \emph {et~al.}(2016)\citenamefont
  {{De Oliveira}}, \citenamefont {Alves},\ and\ \citenamefont
  {Ferreira}}]{VD2}%
  \BibitemOpen
  \bibfield  {author} {\bibinfo {author} {\bibfnamefont {M.~M.}\ \bibnamefont
  {{De Oliveira}}}, \bibinfo {author} {\bibfnamefont {S.}~\bibnamefont
  {Alves}}, \ and\ \bibinfo {author} {\bibfnamefont {S.}~\bibnamefont
  {Ferreira}},\ }\href {\doibase 10.1103/PhysRevE.93.012110} {\bibfield
  {journal} {\bibinfo  {journal} {Physical Review E}\ }\textbf {\bibinfo
  {volume} {93}} (\bibinfo {year} {2016}),\
  10.1103/PhysRevE.93.012110}\BibitemShut {NoStop}%
\bibitem [{\citenamefont {Luck}(1993)}]{luck93}%
  \BibitemOpen
  \bibfield  {author} {\bibinfo {author} {\bibfnamefont {J.~M.}\ \bibnamefont
  {Luck}},\ }\href@noop {} {\bibfield  {journal} {\bibinfo  {journal}
  {Europhys. Lett.}\ }\textbf {\bibinfo {volume} {24}},\ \bibinfo {pages} {359}
  (\bibinfo {year} {1993})}\BibitemShut {NoStop}%
\bibitem [{\citenamefont {Barghathi}\ and\ \citenamefont
  {Vojta}(2014)}]{vojta14c}%
  \BibitemOpen
  \bibfield  {author} {\bibinfo {author} {\bibfnamefont {H.}~\bibnamefont
  {Barghathi}}\ and\ \bibinfo {author} {\bibfnamefont {T.}~\bibnamefont
  {Vojta}},\ }\href@noop {} {\bibfield  {journal} {\bibinfo  {journal} {Phys.
  Rev. Lett.}\ }\textbf {\bibinfo {volume} {113}},\ \bibinfo {pages} {120602}
  (\bibinfo {year} {2014})}\BibitemShut {NoStop}%
\bibitem [{\citenamefont {Schrauth}\ \emph {et~al.}(2018)\citenamefont
  {Schrauth}, \citenamefont {Portela},\ and\ \citenamefont
  {Goth}}]{Schrauth2018}%
  \BibitemOpen
  \bibfield  {author} {\bibinfo {author} {\bibfnamefont {M.}~\bibnamefont
  {Schrauth}}, \bibinfo {author} {\bibfnamefont {J.~S.}\ \bibnamefont
  {Portela}}, \ and\ \bibinfo {author} {\bibfnamefont {F.}~\bibnamefont
  {Goth}},\ }\href {\doibase 10.1103/PhysRevLett.121.100601} {\bibfield
  {journal} {\bibinfo  {journal} {Physical Review Letters}\ }\textbf {\bibinfo
  {volume} {121}},\ \bibinfo {pages} {100601} (\bibinfo {year} {2018})},\
  \Eprint {http://arxiv.org/abs/1803.08753} {arXiv:1803.08753} \BibitemShut
  {NoStop}%
\bibitem [{\citenamefont {Schrauth}\ and\ \citenamefont
  {Portela}(2019{\natexlab{a}})}]{Schrauth2019}%
  \BibitemOpen
  \bibfield  {author} {\bibinfo {author} {\bibfnamefont {M.}~\bibnamefont
  {Schrauth}}\ and\ \bibinfo {author} {\bibfnamefont {J.~S.~E.}\ \bibnamefont
  {Portela}},\ }\href {\doibase 10.1103/PhysRevE.100.062118} {\bibfield
  {journal} {\bibinfo  {journal} {Phys. Rev. E}\ }\textbf {\bibinfo {volume}
  {100}},\ \bibinfo {pages} {062118} (\bibinfo {year}
  {2019}{\natexlab{a}})}\BibitemShut {NoStop}%
\bibitem [{\citenamefont {Schrauth}\ and\ \citenamefont
  {Portela}(2019{\natexlab{b}})}]{Schrauth2019b}%
  \BibitemOpen
  \bibfield  {author} {\bibinfo {author} {\bibfnamefont {M.}~\bibnamefont
  {Schrauth}}\ and\ \bibinfo {author} {\bibfnamefont {J.~S.~E.}\ \bibnamefont
  {Portela}},\ }\href {\doibase 10.1103/PhysRevResearch.1.033061} {\bibfield
  {journal} {\bibinfo  {journal} {Phys. Rev. Research}\ }\textbf {\bibinfo
  {volume} {1}},\ \bibinfo {pages} {033061} (\bibinfo {year}
  {2019}{\natexlab{b}})}\BibitemShut {NoStop}%
\bibitem [{\citenamefont {Hassan}\ \emph {et~al.}(2010)\citenamefont {Hassan},
  \citenamefont {Hassan},\ and\ \citenamefont {Pavel}}]{Hassan2010}%
  \BibitemOpen
  \bibfield  {author} {\bibinfo {author} {\bibfnamefont {M.~K.}\ \bibnamefont
  {Hassan}}, \bibinfo {author} {\bibfnamefont {M.~Z.}\ \bibnamefont {Hassan}},
  \ and\ \bibinfo {author} {\bibfnamefont {N.~I.}\ \bibnamefont {Pavel}},\
  }\href {\doibase 10.1088/1367-2630/12/9/093045} {\bibfield  {journal}
  {\bibinfo  {journal} {New Journal of Physics}\ }\textbf {\bibinfo {volume}
  {12}} (\bibinfo {year} {2010}),\ 10.1088/1367-2630/12/9/093045}\BibitemShut
  {NoStop}%
\bibitem [{\citenamefont {Dayeen}\ and\ \citenamefont
  {Hassan}(2016)}]{Dayeen2016}%
  \BibitemOpen
  \bibfield  {author} {\bibinfo {author} {\bibfnamefont {F.~R.}\ \bibnamefont
  {Dayeen}}\ and\ \bibinfo {author} {\bibfnamefont {M.~K.}\ \bibnamefont
  {Hassan}},\ }\href {\doibase 10.1016/j.chaos.2016.06.006} {\bibfield
  {journal} {\bibinfo  {journal} {Chaos, Solitons and Fractals}\ }\textbf
  {\bibinfo {volume} {91}},\ \bibinfo {pages} {228} (\bibinfo {year}
  {2016})}\BibitemShut {NoStop}%
\bibitem [{\citenamefont {Nagel}\ and\ \citenamefont {Weiss}(2005)}]{crack}%
  \BibitemOpen
  \bibfield  {author} {\bibinfo {author} {\bibfnamefont {W.}~\bibnamefont
  {Nagel}}\ and\ \bibinfo {author} {\bibfnamefont {V.}~\bibnamefont {Weiss}},\
  }\href {\doibase 10.1239/aap/1134587744} {\bibfield  {journal} {\bibinfo
  {journal} {Advances in Applied Probability}\ }\textbf {\bibinfo {volume}
  {37}},\ \bibinfo {pages} {859–883} (\bibinfo {year} {2005})}\BibitemShut
  {NoStop}%
\bibitem [{\citenamefont {Courtat}\ \emph {et~al.}(2011)\citenamefont
  {Courtat}, \citenamefont {Gloaguen},\ and\ \citenamefont {Douady}}]{city}%
  \BibitemOpen
  \bibfield  {author} {\bibinfo {author} {\bibfnamefont {T.}~\bibnamefont
  {Courtat}}, \bibinfo {author} {\bibfnamefont {C.}~\bibnamefont {Gloaguen}}, \
  and\ \bibinfo {author} {\bibfnamefont {S.}~\bibnamefont {Douady}},\ }\href
  {\doibase 10.1103/PhysRevE.83.036106} {\bibfield  {journal} {\bibinfo
  {journal} {Phys. Rev. E}\ }\textbf {\bibinfo {volume} {83}},\ \bibinfo
  {pages} {036106} (\bibinfo {year} {2011})}\BibitemShut {NoStop}%
\bibitem [{\citenamefont {Hassan}\ and\ \citenamefont
  {Rahman}(2016)}]{Hassan2016}%
  \BibitemOpen
  \bibfield  {author} {\bibinfo {author} {\bibfnamefont {M.~K.}\ \bibnamefont
  {Hassan}}\ and\ \bibinfo {author} {\bibfnamefont {M.~M.}\ \bibnamefont
  {Rahman}},\ }\href {\doibase 10.1103/PhysRevE.94.042109} {\bibfield
  {journal} {\bibinfo  {journal} {Physical Review E}\ }\textbf {\bibinfo
  {volume} {94}} (\bibinfo {year} {2016}),\ 10.1103/PhysRevE.94.042109},\
  \Eprint {http://arxiv.org/abs/1604.08699} {arXiv:1604.08699} \BibitemShut
  {NoStop}%
\bibitem [{\citenamefont {Hsu}\ and\ \citenamefont {Huang}(1999)}]{Hsu}%
  \BibitemOpen
  \bibfield  {author} {\bibinfo {author} {\bibfnamefont {H.-P.}\ \bibnamefont
  {Hsu}}\ and\ \bibinfo {author} {\bibfnamefont {M.-C.}\ \bibnamefont
  {Huang}},\ }\href {\doibase 10.1103/PhysRevE.60.6361} {\bibfield  {journal}
  {\bibinfo  {journal} {Phys. Rev. E}\ }\textbf {\bibinfo {volume} {60}},\
  \bibinfo {pages} {6361} (\bibinfo {year} {1999})}\BibitemShut {NoStop}%
\bibitem [{\citenamefont {Liu}\ and\ \citenamefont {Guan}(2018)}]{Liu}%
  \BibitemOpen
  \bibfield  {author} {\bibinfo {author} {\bibfnamefont {W.~Z.}\ \bibnamefont
  {Liu}, \bibfnamefont {XS.}}\ and\ \bibinfo {author} {\bibfnamefont
  {J.}~\bibnamefont {Guan}},\ }\href {\doibase 10.1140/epjb/e2018-90092-x}
  {\bibfield  {journal} {\bibinfo  {journal} {Eur. Phys. J. B}\ }\textbf
  {\bibinfo {volume} {91}},\ \bibinfo {pages} {220} (\bibinfo {year}
  {2018})}\BibitemShut {NoStop}%
\bibitem [{\citenamefont {Lima}\ \emph {et~al.}(2000)\citenamefont {Lima},
  \citenamefont {Moreira}, \citenamefont {Andrade},\ and\ \citenamefont
  {Costa}}]{Lima2000}%
  \BibitemOpen
  \bibfield  {author} {\bibinfo {author} {\bibfnamefont {F.}~\bibnamefont
  {Lima}}, \bibinfo {author} {\bibfnamefont {J.}~\bibnamefont {Moreira}},
  \bibinfo {author} {\bibfnamefont {J.}~\bibnamefont {Andrade}}, \ and\
  \bibinfo {author} {\bibfnamefont {U.}~\bibnamefont {Costa}},\ }\href
  {\doibase https://doi.org/10.1016/S0378-4371(00)00134-5} {\bibfield
  {journal} {\bibinfo  {journal} {Physica A: Statistical Mechanics and its
  Applications}\ }\textbf {\bibinfo {volume} {283}},\ \bibinfo {pages} {100}
  (\bibinfo {year} {2000})}\BibitemShut {NoStop}%
\bibitem [{\citenamefont {{de Oliveira}}\ and\ \citenamefont
  {Dickman}(2005)}]{qssimPRE}%
  \BibitemOpen
  \bibfield  {author} {\bibinfo {author} {\bibfnamefont {M.~M.}\ \bibnamefont
  {{de Oliveira}}}\ and\ \bibinfo {author} {\bibfnamefont {R.}~\bibnamefont
  {Dickman}},\ }\href {\doibase 10.1103/PhysRevE.71.016129} {\bibfield
  {journal} {\bibinfo  {journal} {Phys. Rev. E}\ }\textbf {\bibinfo {volume}
  {71}},\ \bibinfo {pages} {016129} (\bibinfo {year} {2005})}\BibitemShut
  {NoStop}%
\bibitem [{\citenamefont {{de Oliveira}}\ and\ \citenamefont
  {Dickman}(2006)}]{bjp}%
  \BibitemOpen
  \bibfield  {author} {\bibinfo {author} {\bibfnamefont {M.~M.}\ \bibnamefont
  {{de Oliveira}}}\ and\ \bibinfo {author} {\bibfnamefont {R.}~\bibnamefont
  {Dickman}},\ }\href@noop {} {\bibfield  {journal} {\bibinfo  {journal} {Braz.
  J. Phys.}\ }\textbf {\bibinfo {volume} {36}},\ \bibinfo {pages} {685 }
  (\bibinfo {year} {2006})}\BibitemShut {NoStop}%
\bibitem [{\citenamefont {Ferreira}\ \emph {et~al.}(2011)\citenamefont
  {Ferreira}, \citenamefont {Ferreira}, \citenamefont {Castellano},\ and\
  \citenamefont {Pastor-Satorras}}]{silvioQS}%
  \BibitemOpen
  \bibfield  {author} {\bibinfo {author} {\bibfnamefont {S.~C.}\ \bibnamefont
  {Ferreira}}, \bibinfo {author} {\bibfnamefont {R.~S.}\ \bibnamefont
  {Ferreira}}, \bibinfo {author} {\bibfnamefont {C.}~\bibnamefont
  {Castellano}}, \ and\ \bibinfo {author} {\bibfnamefont {R.}~\bibnamefont
  {Pastor-Satorras}},\ }\href {\doibase 10.1103/PhysRevE.84.066102} {\bibfield
  {journal} {\bibinfo  {journal} {Phys. Rev. E}\ }\textbf {\bibinfo {volume}
  {84}},\ \bibinfo {pages} {066102} (\bibinfo {year} {2011})}\BibitemShut
  {NoStop}%
\end{thebibliography}%

\end{document}